\documentclass[pra, twocolumn]{revtex4}

\usepackage{amsmath,amssymb}  
\usepackage{psfrag}  
\usepackage[dvips]{epsfig}  
\usepackage{epsfig}  

\def\R{\mathbb{R}}  
\def\k{\mathfrak{k}}  
\def\su{\text{su}} 
\def\so{\text{so}}  
\def\SU{\text{SU}}

\begin{document}   

\title{Optimal generation of single-qubit operation from an always-on
  interaction\\ by algebraic decoupling}

\author{Jun Zhang$^{1, 2}$ and K. Birgitta Whaley$^{1}$}
\affiliation{$^1$Department of Chemistry and Pitzer Center for
  Theoretical Chemistry,
  University of California, Berkeley, CA 94720\\
  $^2$Department of Electrical Engineering and Computer Sciences,
  University of California, Berkeley, CA 94720} 

\date{\today}
  
\begin{abstract}  
  We present a direct algebraic decoupling approach to generate
  arbitrary single-qubit operations in the presence of a constant
  interaction by applying local control signals. To overcome the
  difficulty of undesirable entanglement generated by the untunable
  interaction, we derive local control fields that are designed to
  both drive the qubit systems back to unentangled states at the end
  of the time interval over which the desired single-qubit operation
  is completed.  This approach is seen to be particularly relevant for
  the physical implementation of solid-state quantum computation and
  for the design of low-power pulses in NMR.
\end{abstract}   
  
\maketitle 

Most schemes for implementation of quantum computation require
achieving both single- and two-qubit operations in order to realize
the speed-up associated with quantum algorithms~\cite{Barenco:95}.
Typically, single-qubit operations are implemented by external local
control fields applied to individual qubits, whereas two-qubit
operations are generated by interplay of the couplings between qubits
and local fields~\cite{Zhang:04a}. To simplify engineering design and
reduce decoherence channels, many physical proposals of quantum
computation have ``always-on'' and fixed couplings between qubits.
This is often the case in arrays of solid-state qubits, which
constitute a very attractive research direction because of the
inherent scalability of the required microfabrication techniques. For
example, in superconducting systems the interaction between qubits is
often coupled by a capacitor or inductor, whose value is fixed by the
fabrication and generally cannot be tuned during the
computations~\cite{Strauch:03, Yamamoto:03}. Several variable coupling
schemes have been suggested~\cite{Mooij:99, Makhlin:01, Clarke:02,
  Averin:03, Britton:04}, but it is generally agreed that none of
these is completely satisfactory.
In contrast, proposals for using electron spins in quantum dots do in
principle allow for electrical gating of the exchange spin-spin
interaction~\cite{Loss:98,Burkard:99}, but engineering such control in
practice still remains extremely challenging (See, e.g.,
Ref.~\cite{DasSarma:05} and references therein.)  Other spin-coupled
implementations of current interest include endohedral fullerenes for
which non-tunable magnetic dipolar coupling between electron spins of
neighboring endohedral species provide the required qubit
coupling~\cite{Suter:02,Twamley:03}.

While coupling between qubits is essential for the implementation of
two-qubit operations, an always-on and untunable coupling leads to
additional complications in implementation of single-qubit operations,
because the qubits may become entangled.  This general issue is also
encountered in nuclear magnetic resonance. For the short and high
power ``hard'' pulses, sophisticated refocusing schemes have been
developed to decouple the subsystems~\cite{Freeman:98}. However, for
the low-power pulses often used in homonuclear spin systems, the radio
frequency signals can be of the same order of magnitude as the
coupling strengths~\cite{Kupce:95, Lieven:04}. In this ``soft'' pulse
situation, the single-qubit rotations cannot be assumed to be
implemented instantaneously as is the case with hard pulses, and
consequently the interactions may strongly affect the intended
operations.  To overcome this general problem in the context of
quantum information processing, Ref.~\cite{Lidar:02a} employed an
encoding of logical qubits, while logic operations are performed using
an analogue of the NMR selective recoupling method.
Ref.~\cite{Zhou:02} presented another encoding scheme to realize
universal quantum computation on carefully designed interaction free
subspaces.  Recently, Ref.~\cite{Benjamin:03} showed that one can also
avoid this undesirable entanglement by tuning the transition energies
of individual qubits.
More complex schemes involving auxiliary degrees of freedom have also
been developed~\cite{Suter:02,Twamley:03}.  Here we present a {\em
  direct} approach for the design of local control signals in the
presence of an always-on interaction.  To eliminate the accompanying
entanglement generated by this untunable interaction, we derive local
control fields that drive the qubit systems back to unentangled states
at the end of the time interval over which the desired single-qubit
operation is completed.

We consider a two-qubit system with the following Hamiltonian:
\begin{eqnarray}
  \label{eq:1}
H&=&\frac{\omega_1}2(\cos\phi_1 \sigma_x^1+\sin\phi_1 \sigma_y^1)\nonumber\\
&& +\frac{\omega_2}2(\cos\phi_2 \sigma_x^2+\sin\phi_2 \sigma_y^2)+
\frac{J}2\sigma_z^1\sigma_z^2,
\end{eqnarray}
where $\sigma_x$, $\sigma_y$, and $\sigma_z$ are the Pauli matrices,
$J$ the always-on and untunable coupling strength, $\omega_j$ and
$\phi_j$ the amplitude and phase of the external control fields,
respectively. This Hamiltonian describes a wide range of two-qubit
systems, e.g., solid-state quantum computation with superconducting
circuits as proposed in~\cite{Makhlin:01}, as well as NMR
double-resonance $J$ cross polarization with two RF fields in the XY
plane of the doubly rotating frame~\cite{Maj:95}.  Note that the
Hamiltonian in Eq.~\eqref{eq:1} can be transformed by local unitary
operations into a Hamiltonian with YY coupling and $\sigma_x$,
$\sigma_z$ local terms, or into one with XX coupling and $\sigma_y$,
$\sigma_z$ local terms. Therefore, the results in this paper are also
applicable to these two types of Hamiltonians.  The spin coupling $J$
here can be neglected only for short and high-power pulses, since for
low-power pulses $\omega_1$ and $\omega_2$ can be of the same order as
$J$. We now show how to design amplitude-modulated pulses that can
generate an arbitrary single-qubit operation for such a system. The
control pulses are derived by algebraically decoupling the two-qubit
Hamiltonian into two unentangled single-qubit systems. This approach
is easy to implement in the physical systems of interest and avoids
the encoding overheads associated with the encoding schemes of
Refs.~\cite{Lidar:02a, Zhou:02}.

An arbitrary local unitary operation $k_1$ on two qubits, the target
operation here, can be written using Euler's XYX decomposition as:
\begin{eqnarray}
  \label{eq:4}
k_1&=&(e^{-i \alpha_1 \sigma_x/2}\otimes e^{-i \alpha_2 \sigma_x/2})
 (e^{-i\beta_1 \sigma_y/2}\otimes e^{-i\beta_2 \sigma_y/2})\nonumber \\
&& \times (e^{-i\gamma_1 \sigma_x/2} \otimes  e^{-i\gamma_2 \sigma_x/2}),
\end{eqnarray}
where $\alpha_j$, $\beta_j$, and $\gamma_j$ are the Euler angles. In
order to generate $k_1$, we need to generate arbitrary $\sigma_x$ and
$\sigma_y$ rotations on each qubit. However, we can simplify this
problem to the generation of only $\sigma_x$ rotations
$k_2=e^{-i\gamma_1 \sigma_x/2} \otimes e^{-i\gamma_2 \sigma_x/2}$ from
the following Hamiltonian:
\begin{eqnarray}
  \label{eq:3}
  H_1=\frac{\omega_1}2 \sigma_x^1
+\frac{\omega_2}2 \sigma_x^2+ \frac{J}2\sigma_z^1\sigma_z^2,
\end{eqnarray}
where $H_1$ is obtained from Eq.~\eqref{eq:1} by setting the phases
$\phi_1=\phi_2=0$ for external control pulses.  This leads immediately
to the implementations of the first and third operations in
Eq.~\eqref{eq:4}. For the second operation in Eq.~\eqref{eq:4}, we can
set $\phi_1=\phi_2=\pi/2$ in Eq.~\eqref{eq:1} to obtain a Hamiltonian
\begin{equation}
  \label{eq:66}
H_2=\frac{\omega_1}2 \sigma_y^1
+\frac{\omega_2}2 \sigma_y^2+ \frac{J}2\sigma_z^1\sigma_z^2=V^\dag H_1 V,
\end{equation}
where $V=e^{i\pi/2 \sigma_z/2}\otimes e^{i \pi/2\sigma_z/2}$. Now if
$H_1$ generates a quantum operation $e^{-i\beta_1 \sigma_x/2}\otimes
e^{-i\beta_2 \sigma_x/2}$, $H_2$ will generate the second operation in
Eq.~\eqref{eq:4}, since
\begin{eqnarray*}
e^{-i\beta_1 \sigma_y/2}\otimes e^{-i\beta_2 \sigma_y/2}=
V^\dag(e^{-i\beta_1 \sigma_x/2}\otimes e^{-i\beta_2 \sigma_x/2})V.
\end{eqnarray*}
Our task has therefore been simplified to the generation of an
arbitrary local unitary operation $e^{-i\gamma_1 \sigma_x/2} \otimes
e^{-i\gamma_2 \sigma_x/2}$ from the Hamiltonian $H_1$ in
Eq.~\eqref{eq:3}. We now observe that the terms $i\sigma_x^1/2$,
$i\sigma_x^2/2$, and $i\sigma_z^1\sigma_z^2/2$ appearing in $H_1$
generate the following Lie algebra:
\begin{eqnarray}
  \label{eq:5} 
  \k_1=\frac{i}2\{\sigma_x^1,\ \sigma_x^2,\  \sigma_z^1\sigma_y^2,\
  \sigma_y^1\sigma_z^2,\  \sigma_y^1\sigma_y^2,\ \sigma_z^1\sigma_z^2\}. 
\end{eqnarray}
It is straightforward to show that $\k_1$ satisfies the same
commutation relations as $\so(4)$, where $\so(4)$ denotes the Lie
algebra formed by all the $4\times 4$ real skew symmetric matrices.
Therefore, $\k_1$ is isomorphic to $\so(4)$. We also know that
$\so(4)$ is isomorphic to $\su(2)\otimes \su(2)$~\cite{Cornwell:97}.
To realize this fact, let
\begin{eqnarray}
  \label{eq:6}
  \epsilon_x^1= \frac{\sigma_x^1-\sigma_x^2}4,\  
\epsilon_y^1=\frac{\sigma_y^1\sigma_y^2+\sigma_z^1\sigma_z^2}4,
  \epsilon_z^1=\frac{\sigma_z^1\sigma_y^2-\sigma_y^1\sigma_z^2}4,\nonumber\\
  \epsilon_x^2=\frac{\sigma_x^1+\sigma_x^2}4,\
  \epsilon_y^2=\frac{\sigma_y^1\sigma_y^2-\sigma_z^1\sigma_z^2}4,
  \epsilon_z^2=\frac{\sigma_z^1\sigma_y^2+\sigma_y^1\sigma_z^2}4,\nonumber\\
\end{eqnarray}
and use $\k_2$ to denote the Lie algebra generated by
$\{i\epsilon_x^1$, $i\epsilon_y^1$, $i\epsilon_z^1$, $i\epsilon_x^2$,
$i\epsilon_y^2$, $i\epsilon_z^2\}$.  We have the following commutation
relations for $\k_2$:
\begin{eqnarray*}
\begin{array}{c|cccccc}
[\cdot, \cdot]&i\epsilon_x^1&i\epsilon_y^1&i\epsilon_z^1
&i\epsilon_x^2&i\epsilon_y^2&i\epsilon_z^2\\ \hline
i\epsilon_x^1&0&-i\epsilon_z^1&i\epsilon_y^1&0&0&0\\ 
i\epsilon_y^1&i\epsilon_z^1&0&-i\epsilon_x^1&0&0&0\\
i\epsilon_z^1&-i\epsilon_y^1&i\epsilon_x^1&0&0&0&0\\
i\epsilon_x^2&0&0&0&0&-i\epsilon_z^2&i\epsilon_y^2\\
i\epsilon_y^2&0&0&0&i\epsilon_z^2&0&-i\epsilon_x^2\\
i\epsilon_z^2&0&0&0&-i\epsilon_y^2&i\epsilon_x^2&0
\end{array}
\end{eqnarray*}
It is clear that the Lie algebra $\k_2$ satisfies the same commutation
relations as $i/2\{\sigma_x^1$, $\sigma_y^1$, $\sigma_z^1$,
$\sigma_x^2$, $\sigma_y^2$, $\sigma_z^2\}$, and therefore it is
isomorphic to $\su(2)\otimes \su(2)$. This isomorphism allows great
simplification for the generation of single-qubit operation from
Hamiltonian~\eqref{eq:3}, because it provides an algebraic way to
decouple the entangling Hamiltonian into two unentangled single-qubit
Hamiltonians. To our knowledge, this fact has not been recognized,
although a transformation similar to~\eqref{eq:6} was presented
in~\cite{Maj:95}.

We can now rewrite $H_1$ as
\begin{eqnarray}
  \label{eq:7}
  H_1=(\omega_1-\omega_2)\epsilon_x^1+J\epsilon_y^1
+(\omega_1+\omega_2)\epsilon_x^2-J\epsilon_y^2,
\end{eqnarray}
and the desired generator $k_2$ as
\begin{eqnarray}
  \label{eq:8}
k_2= e^{-i\gamma_1 \sigma_x/2} \otimes e^{-i\gamma_2 \sigma_x/2}
=e^{-i((\gamma_1-\gamma_2)\epsilon_x^1+(\gamma_1+\gamma_2)\epsilon_x^2)}.
\end{eqnarray}
We now transform the parameters in $k_2$ and $H_1$ to reformulate the
problem as generation of the local unitary
\begin{eqnarray}
 \label{eq:50}
  e^{-i((\gamma_1-\gamma_2)\sigma_x^1/2 +
  (\gamma_1+\gamma_2)\sigma_x^2/2)}
\end{eqnarray}
from the Hamiltonian
 \begin{eqnarray}
 \label{eq:51}
\frac{\omega_1-\omega_2}2\sigma_x^1+\frac{J}2\sigma_y^1
+\frac{\omega_1+\omega_2}2\sigma_x^2-\frac{J}2\sigma_y^2.
 \end{eqnarray}
 To obtain Eqs.~\eqref{eq:50} and~\eqref{eq:51} we simply replaced
 $\epsilon_\alpha^j$ in Eqs.~\eqref{eq:7} and~\eqref{eq:8} by
 $\sigma_\alpha^j/2$, which is warranted by the fact that
 $\{i\epsilon_x^1, i\epsilon_y^1, i\epsilon_z^1, i\epsilon_x^2,
 i\epsilon_y^2, i\epsilon_z^2\}$ is isomorphic to the Lie algebra
 $\su(2)\otimes \su(2)$. Taking advantage of the commutation of
 $\sigma_\alpha^1$ and $\sigma_\beta^2$, the problem then naturally
 decomposes into two well-defined problems of generating single-qubit
 rotations:

(1) Generate $e^{-i(\gamma_1-\gamma_2)\sigma_x^1/2}$ from the
Hamiltonian $(\omega_1-\omega_2)\sigma_x^1/2+ J\sigma_y^1/2$; and

(2) Generate $e^{-i(\gamma_1+\gamma_2)\sigma_x^2/2}$ from the
Hamiltonian $(\omega_1+\omega_2)\sigma_x^2/2-J\sigma_y^2/2$.

By making the transformation from $\k_1$ to $\su(2)\otimes\su(2)$, we
have therefore arrived at two decoupled single-qubit quantum systems,
solutions of which both reduce to a general steering problem on the Lie
group $\SU(2)$ with dynamics determined by the Schr\"{o}dinger
equation:
\begin{equation}
 \label{eq:12}
   i\dot U= (\frac{\omega(t)}2\sigma_x+\frac{J}2\sigma_y) U,\quad U(0)=I.
\end{equation}
In general, $\omega(t)$ is a time dependent external control field.
The minimum energy control on the Lie group $\SU(2)$ has been studied
in Ref.~\cite{DAl01}. Here we will use Lie-Poisson reduction to
derive both the minimum energy and time optimal control. We will also
give a good approximate solution to this problem.

Pontryagin's Maximal Principle provides an important mathematical tool
for solving the optimal steering problem on Lie groups~\cite{Pon:62}.
We solve for a control field $\omega$ that drives the system from the
initial operation $U(0)=I$ to a prescribed target operation
$U(T)=e^{-i\gamma/2 \sigma_x}$, which also minimizes the cost function:
\begin{equation}
\label{eq:29}
  J=\int_0^T L(\omega(t))\, dt,
\end{equation}
where $L(\omega(t))$ is a general functional of the control field
$\omega(t)$ and is often referred as the running cost. The first step
is to construct the control Hamiltonian:
\begin{eqnarray}
\label{eq:30}
 {\cal H}(t)=L(\omega(t)) +\langle M, -i {U(t)}^\dag
(\frac{{\omega(t)}}2\sigma_x+\frac{J}2\sigma_y){U(t)}\rangle ,
\end{eqnarray}
where $M$ is a constant matrix in $\su(2)$, and $\langle X,
Y\rangle=\text{Tr} (XY^\dag)$ is an inner product defined on $\su(2)$.
For the ease of notation, we will suppress the time parameter unless
otherwise specified.  Proceeding further, we can write
Eq.~\eqref{eq:30} as
\begin{eqnarray}
\label{eq:31}
{\cal H}=L(\omega) -\frac12\langle M, i {U}^\dag{\sigma_x}
 {U}\rangle{\omega}- \frac12\langle M, i{U}^\dag{\sigma_y} {U}\rangle  J.
\end{eqnarray}
We then let $p_1=\langle M, i {U}^\dag{\sigma_x}{U}\rangle/2$,
$p_2=\langle M, i {U}^\dag{\sigma_y}{U}\rangle/2$, and $p_3=\langle M,
i {U}^\dag{\sigma_z}{U}\rangle/2$. Taking the derivative of $p_1$, we
obtain
\begin{eqnarray}
\label{eq:34}
  \dot p_1=\frac12\langle M, i\dot {U}^\dag \sigma_x {U}\rangle
+\frac12\langle M, i{U}^\dag \sigma_x \dot{U}\rangle
=Jp_3.
\end{eqnarray}
By similar means, we can get the dynamics for $p_2$ and $p_3$.
Combining these together, we have the dynamics for $p_j$ as
\begin{eqnarray}
\dot p_1&=&Jp_3\nonumber\\
\label{eq:36}
\dot  p_2&=&-\omega p_3,\\
\dot p_3&=&\omega p_2-Jp_1\nonumber.
\end{eqnarray}
Eq.~\eqref{eq:36} is indeed the Lie-Poisson equation on
$\SU(2)$~\cite{Marsden:99}. It is easy to verify that the functions
$C_1=p_1^2+p_2^2+p_3^2$ and $C_2=p_1^2+2Jp_2$ are both invariant along
the system trajectory. For a general optimal control problem on Lie
group $\SU(2)$, one usually needs to solve a set of six differential
equations.  However, here we can obtain the reduced dynamics of three
differential equations in Eq.~\eqref{eq:36} by using the Lie-Poisson
reduction theorem~\cite{Marsden:99}. The two conditions for this
theorem are the vector field is right invariant and the cost function
being independent of the quantum states, and they are both satisfied
in this case.  This reduces the number of the differential equations to
be solved by half. This approach is a general technique applicable
also to analysis of optimal control on multi-qubit systems.

Now we derive the time optimal control, that is, the form of control
field $\omega$ that achieves the target operation in the minimum time
possible. In this case, the cost function can be written as
\begin{equation*}
  \min J=\int_0^T 1 \, dt,
\end{equation*}
and the running cost $L(\omega)=1$ is independent of control field
$\omega$.  The corresponding control Hamiltonian is
\begin{equation}
\label{eq:60}
 {\cal H}=1-\omega p_1-Jp_2.
\end{equation}
Let us assume that the control field $\omega$ is restricted to an
interval $[\omega_m, \omega_M]$ due to physical bounds on the
available frequencies.  From Pontryagin's Maximum Principle, the
optimal control field $\bar{\omega}(t)$ will minimize the control
Hamiltonian~\eqref{eq:60} pointwise. Therefore, $\bar{\omega}(t)$ can
take only extremal values:
\begin{equation*}
  \bar{\omega}(t)=  \begin{cases}
    {\omega}_m,& \text{if }p_1(t)\ge 0,\\
\omega_M,&\text{if }p_1(t) <0.
  \end{cases}
\end{equation*}
This means that the time optimal control will switch back and forth
between two extremal control values. This procedure is usually called
Bang-Bang control in the control theory literature. (Note that the
term ``Bang-Bang control'' has recently been adopted with a somewhat
different meaning in the study of dynamical coupling of open quantum
systems~\cite{Viola:99, Viola:02, Wu:02, Byrd:02}, where it is
referred to performing a set of instantaneously or as fast as
physically possible hard pulses.) In Ref.~\cite{Zhang:04a}, we have
derived a constructive approach to achieve a desired single-qubit
target quantum operation by switching between two such constant control
fields. From the arguments above, we conclude now that this is indeed
a time optimal control strategy.

Next we consider the minimum energy control which is encoded by the
following cost function:
\begin{equation*}
  \min J=\frac12\int_0^T \omega^2(t) dt.
\end{equation*}
The control Hamiltonian is now
\begin{equation*}
   {\cal H}=\frac12 \omega^2-\omega p_1-Jp_2,
\end{equation*}
and thus the minimum energy control is $\bar{\omega}(t)=p_1(t)$.
Taking the derivative of $\dot p_1$ in Eq.~\eqref{eq:36}, we have
\begin{eqnarray}
  \label{eq:37}
  \ddot{\bar{\omega}}=\bigg(\frac{C_2}2-J\bigg)
\bar{\omega}-\frac12 \bar{\omega}^3.
\end{eqnarray}
where $C_2=p_1^2+2Jp_2$ is constant along the optimal trajectory. The
solution of this differential equation is given by the Jacobi elliptic
function Cn~\cite{DAl01}:
\begin{eqnarray}
  \label{eq:38}
  \bar{\omega}(t)=2bk\text{Cn}(bt+f, k),
\end{eqnarray}
where $b$, $f$, and $k$ are real numbers. We can numerically determine
the values of these parameters such that the control
function~\eqref{eq:38} steers the Hamiltonian~\eqref{eq:3} from the
initial operator $U(0)=I$ to the target operation $U(T)=e^{-i\gamma/2
  \sigma_x}$.

Finally we give an approximate solution to the control problem in
terms of a sinusoidal control function $\omega=A\cos(\upsilon t)$, and
show that this can be further fidelity optimized. Letting
$U_1=e^{i\upsilon/2\sigma_y t}U$, we transform the
system~\eqref{eq:12} into the following form:
\begin{align}
i \dot U_1=\{\frac{A}2(1+\cos 2\upsilon t)\frac{\sigma_x}2
+(J-\upsilon)\frac{\sigma_y}{2}+
\frac{A}2\sin 2\upsilon t\frac{\sigma_z}2\}U_1\nonumber \\
 \label{eq:20} 
\end{align}
The usual technique to solve this differential equation in the context
of NMR is to drop the oscillating terms to get a time-independent
Hamiltonian (rotating wave approximation). In order to get a more
accurate solution, we use the Wei-Norman formula~\cite{Wei:64} to
transform Eq.~\eqref{eq:20} to a dynamical system on $\R^3$. From
Euler's ZXZ decomposition, a general solution to Eq.~\eqref{eq:20} can
be written as
\begin{eqnarray}
\label{eq:55}
U_1(t)=e^{-i\alpha_1(t)\sigma_z/2}e^{-i\alpha_2(t)\sigma_x/2}
e^{-i\alpha_3(t)\sigma_z/2}.
\end{eqnarray}
Taking the time derivative of $U_1$, we have
\begin{eqnarray}
&i\dot U_1
=\left\{(\dot\alpha_2 \cos\alpha_1
  +\dot\alpha_3\sin\alpha_1\sin\alpha_2 )\dfrac{\sigma_x}2 \nonumber \right.
+(\dot\alpha_2\sin\alpha_1\nonumber\\
\label{eq:21}
&-\dot\alpha_3\cos\alpha_1\sin\alpha_2)\dfrac{\sigma_y}2
\left.+(\dot\alpha_1+\dot\alpha_3\cos\alpha_2)\dfrac{\sigma_z}2
\right\}U_1.
\end{eqnarray}
Comparing Eqs.~\eqref{eq:20} and~\eqref{eq:21},
 we get
 \begin{eqnarray*}
    \label{eq:22}
    \left[ \begin{matrix}
     A/2(1+\cos 2\upsilon t)\\
     J-\upsilon\\
     A/2\sin 2\upsilon t
    \end{matrix} \right]=
 \left[
   \begin{matrix}
     0&\cos\alpha_1&\sin\alpha_1\sin\alpha_2\\
 0&\sin\alpha_1&-\cos\alpha_1\sin\alpha_2\\
 1&0&\cos\alpha_2
   \end{matrix}\right]
 \left[
   \begin{matrix}
     \dot\alpha_1\\ \dot\alpha_2\\ \dot\alpha_3
   \end{matrix}\right].
 \end{eqnarray*}
 Therefore, we obtain a set of differential equations of the
 parameters $\alpha_j$ in $\R^3$:
\begin{eqnarray}
   \label{eq:23}
\left[
  \begin{matrix}
    \dot\alpha_1\\ \dot\alpha_2\\ \dot\alpha_3
  \end{matrix}\right]=
\left[\begin{matrix}
    -\frac{\sin\alpha_1}{\tan\alpha_2}&\frac{\cos\alpha_1}{\tan\alpha_2}&1\\
\cos\alpha_1&\sin\alpha_1&0\\
\frac{\sin\alpha_1}{\sin\alpha_2}&-\frac{\cos\alpha_1}{\sin\alpha_2}&0 
\end{matrix}\right]
 \left[\begin{matrix}
    A/2(1+\cos 2\upsilon t)\\
    J-\upsilon\\
    A/2\sin 2\upsilon t
   \end{matrix} \right].
\end{eqnarray}
Rigorous solution of Eq.~\eqref{eq:23} cannot be given in terms of
elementary functions. However, when $\upsilon$ is close to $J$, a very
good approximate solution can be found as:
\begin{eqnarray*}
\alpha_1(t)&=&\frac{A}{2\upsilon}\sin^2 \upsilon t,\\
\alpha_2(t)&=&\frac{A}2t+\frac{A}{4J}\sin 2\upsilon t,\\
\alpha_3(t)&=&0.
\end{eqnarray*}
Hence a sinusoidal control function $\omega=A\cos(\upsilon t)$ can
achieve the following unitary operation at a final time $T$:
\begin{eqnarray}
  \label{eq:9}
U(T)=e^{i\upsilon T\sigma_y/2}e^{-i\alpha_1(T)\sigma_z/2}
e^{-i\alpha_2(T)\sigma_x/2}
\end{eqnarray}
To generate an operation $e^{-i\gamma/2\sigma_x}$ with $\gamma\in [0,
2\pi]$, we only need that $\upsilon=J$, $AT/2=\gamma$, and $JT=2n\pi$,
where $n$ is an integer. This leads to the conditions:
\begin{eqnarray}
  \label{eq:25}
 \upsilon=J,\quad A=\frac{\gamma J}{n\pi}, \quad T=\frac{2n\pi}J .
\end{eqnarray}
From Eq.~\eqref{eq:8}, to implement a two-qubit local unitary
operation $ e^{-i\gamma_1 \sigma_x/2} \otimes e^{-i\gamma_2
  \sigma_x/2}$, we require that
\begin{eqnarray}
 \label{eq:26}
\omega_1=\frac{\gamma_1 J}{n\pi}\cos Jt,\quad
\omega_2=\frac{\gamma_2 J}{n\pi}\cos Jt,
\end{eqnarray}
with a time duration $2n\pi/J$. Note that this is an approximate
solution. Numerical simulations reveal that the greater the value of
$n$, the better the approximation.  To improve the accuracy, we can
numerically search the variable parameters that maximize the fidelity
of the actual achieved operation and the desired target operation,
using parameter values given by Eq.~\eqref{eq:25} as a starting guess.
This gives rise to a fidelity optimized control.

\begin{figure}[tb]
\begin{center}
\begin{tabular}{cc}
\scriptsize
\psfrag{A}[][]{$0$}
\psfrag{B}[][]{$10$}
\psfrag{C}[][]{$20$}
\psfrag{D}[][]{$30$}
\psfrag{EEE}[][]{$-80$}
\psfrag{FFF}[][]{$-40$}
\psfrag{G}[][]{$0$}
\psfrag{FF}[][]{$40$}
\psfrag{EE}[][]{$80$}
\psfrag{T}[][]{$T\text{(ms)}$}
\psfrag{Y}[][]{$\omega\text{(Hz)}$}
\includegraphics[width=42mm]{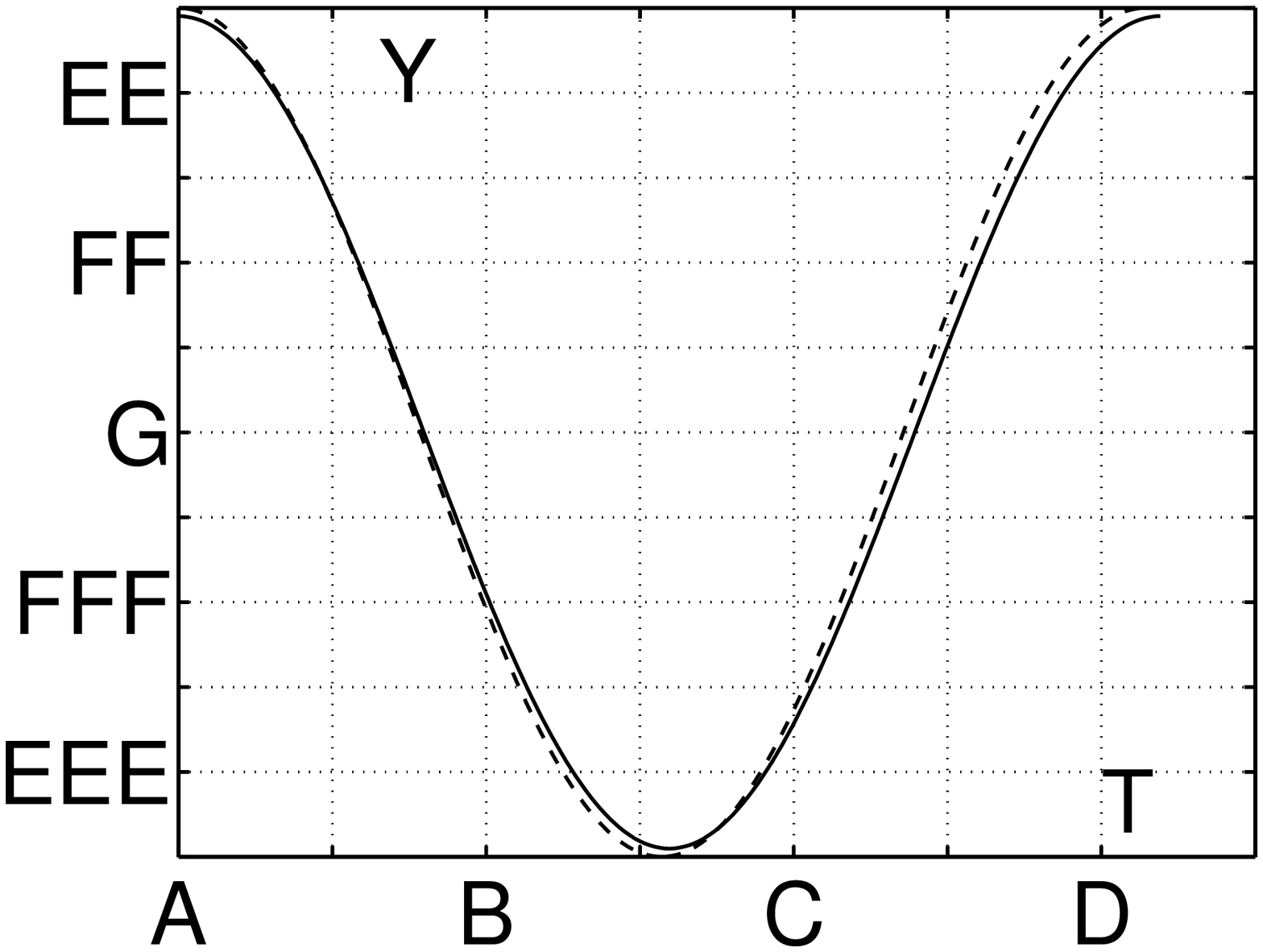}&
\scriptsize
\psfrag{A}[][]{$0$}
\psfrag{B}[][]{$10$}
\psfrag{C}[][]{$20$}
\psfrag{D}[][]{$30$}
\psfrag{EEE}[][]{$-10$}
\psfrag{EE}[][]{$10$}
\psfrag{FF}[][]{$0$}
\psfrag{T}[][]{$T\text{(ms)}$}
\psfrag{Y}[][]{$\omega\text{(Hz)}$}
\includegraphics[width=42mm]{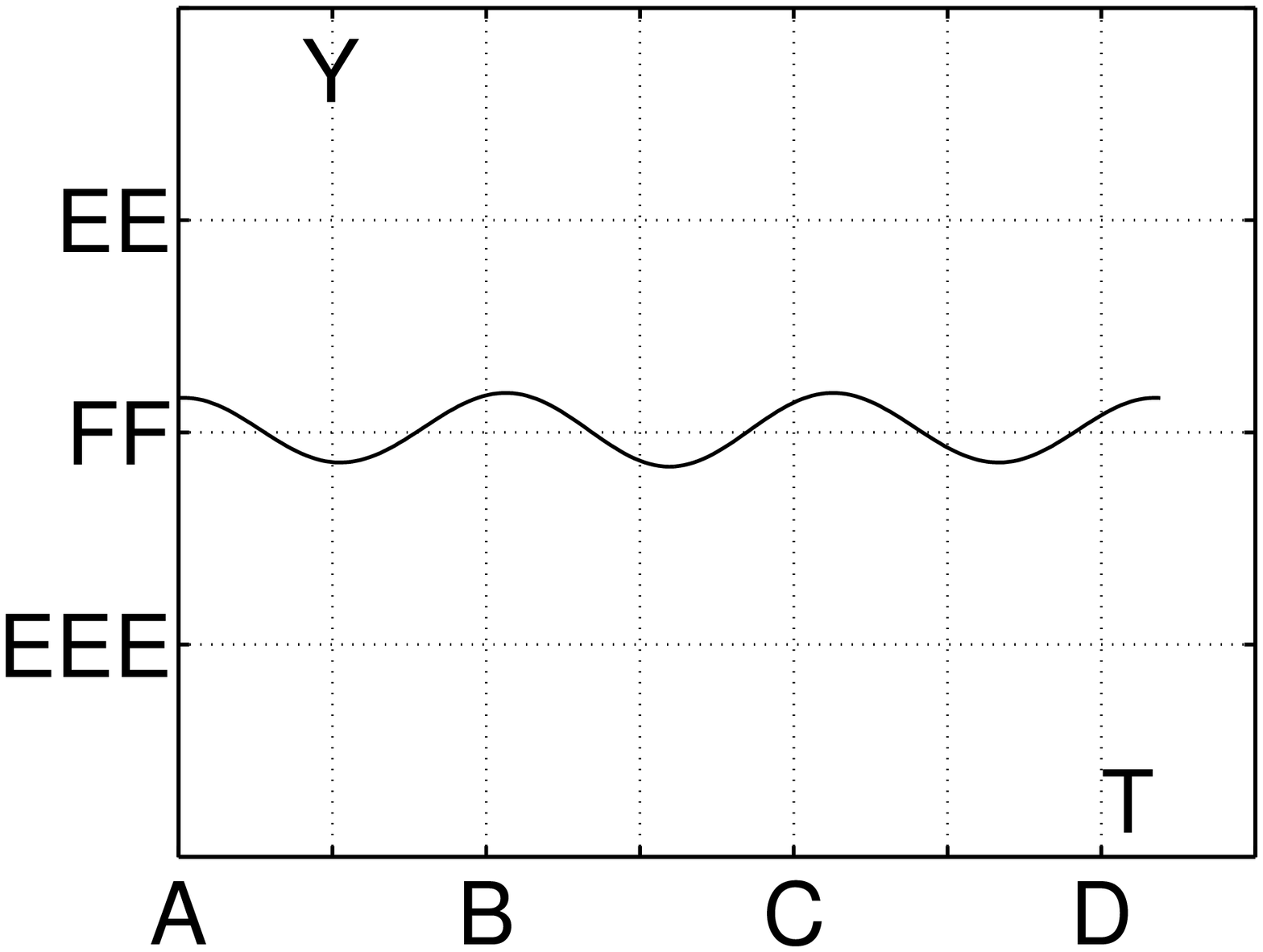}\\
(A)&(B)
\end{tabular}
\end{center}
\caption{Control functions that generate the single spin operation
  $e^{-i\pi/4\sigma_x^1}$ when $J=200$ Hz. (A) Dashed line:
  approximate control in Eq.~\eqref{eq:39}; solid line:
  fidelity optimized control in Eq.~\eqref{eq:41}; (B) the difference
  between fidelity optimized control and minimum energy control in
  Eq.~\eqref{eq:42}.}
\label{fig:1}
\end{figure}

We illustrate the control strategies aforementioned with an example
relevant to NMR experiments on homonuclear spin
systems~\cite{Kupce:95, Lieven:04}.  Let $J=200$ Hz in Eq.~\eqref{eq:1}, and our
target the generation of a $90^\circ$ rotation about $x$-axis on the
first qubit, that is, the operation $e^{-i\pi/4\sigma_x^1}$.  Choosing
$\phi_1=\phi_2=0$, $n=1$, the analytic approximation Eq.~\eqref{eq:25}
yields
\begin{eqnarray}
  \label{eq:40}
 \upsilon=200\, \text{Hz},
\quad A=100\, \text{Hz}, 
\quad T=10\pi\, \text{ms},
\end{eqnarray}
which leads to an approximate solution:
\begin{eqnarray}
 \label{eq:39}
\omega_1=100\cos 200t,\quad \omega_2=0.
\end{eqnarray}
Numerical optimization via the maximization of the fidelity of actual
achieved operation and the desired target operation, using the
parameters in Eq. \eqref{eq:40} as an initial guess, leads to the
following fidelity optimized control function
\begin{eqnarray}
 \label{eq:41}
\omega_1=98.062 \cos 196.900 t,\quad \omega_2=0,
\end{eqnarray}
with a final time $T=31.911$ ms. These two control functions, the
analytic approximation and the fidelity optimized function, are shown
in Fig.~\ref{fig:1}(A). The alternative minimum energy control
function is determined numerically, resulting in parameters
$b=210.744$, $f=-0.00549$, and $k=0.00236$, in Eq.~\eqref{eq:38}, and
the minimum energy control:
\begin{eqnarray}
  \label{eq:42}
  \omega_1=99.678 \text{Cn}(210.744 t-0.00549, 0.00236).
\end{eqnarray}
This minimum energy control solution is seen to be very close to the
fidelity optimized control function of Eq.~\eqref{eq:41}. The
difference between these two control functions is plotted in
Fig.~\ref{fig:1}(B).

In summary, we have presented a design of direct control pulses that
implement any arbitrary single spin operation in a quantum system with
an always-on interaction. This is crucial for the physical
implementation of solid-state quantum computations and for the
low-power pulses design in NMR.  We used an algebraic approach to
decouple the two-qubit system to overcome the difficulty of
undesirable entanglement generated by the untunable interaction.  To
generate the desired target quantum operations, we derived three
control strategies, i.e., minimum time, minimum energy, and an
approximate solution. The first two optimal strategies implement the
desired target quantum operation exactly, whereas the approximate
control can be further optimized for perfect fidelity. The advantage
of the approximate control is that we can easily determine the control
parameters, and it is close to the minimum energy control.

\begin{acknowledgments}
  We thank the NSF for financial support under ITR Grant No.
  EIA-0205641, and DARPA and ONR under Grant No. FDN0014-01-1-0826 of
  the DARPA SPINs program.
\end{acknowledgments}

\bibliographystyle{apsrev} 

\end{document}